\documentclass[cameraready]{Interspeech}

\usepackage{amsmath,amssymb,amsfonts}

\usepackage{graphicx}
\usepackage{booktabs,multirow,makecell,array}
\usepackage{tablefootnote}
\usepackage[table]{xcolor}
\usepackage{pifont}

\usepackage{float}

\usepackage{hyperref}
\hypersetup{hidelinks}

\newcolumntype{C}[1]{>{\centering\arraybackslash}m{#1}}

\definecolor{rk}{rgb}{0.3, 0.0, 0.9}

\usepackage{soul}
\newcommand{\hly}[2][yellow!50]{%
  \begingroup\setlength\fboxsep{1pt}\colorbox{#1}{\strut #2}\endgroup
}

\title{Distilling LLM Semantic Priors into Encoder-Only Multi-Talker ASR with Talker-Count Routing}

\author[]{Hao}{Shi}
\author[]{Yusuke}{Fujita}
\author[]{Roman}{Koshkin}
\author[]{Mengjie}{Zhao}
\author[]{Yuan}{Gao}
\author[]{Lianbo}{Liu}
\author[]{Yui}{Sudo}

\address{
SB Intuitions, Tokyo, Japan
}

\email{hshi@ieee.org}


\AtBeginDocument{%
  \setlength{\abovedisplayskip}{0pt}%
  \setlength{\belowdisplayskip}{0pt}%
  \setlength{\abovedisplayshortskip}{0pt}%
  \setlength{\belowdisplayshortskip}{0pt}%
}

\begin{document}

\maketitle

\begin{abstract}
Large language models (LLMs) provide strong semantic priors that can improve multi-talker automatic speech recognition (MT-ASR), but using an LLM as an autoregressive decoder is computationally expensive and remains fragile under heavy overlap. 
In this paper, we propose an \emph{encoder-only} MT-ASR framework that \emph{adapts} an LLM to multi-talker conditioning and \emph{distills} its semantic guidance into the encoder during training, while retaining fast CTC-style decoding at inference. 
Our model employs a post-encoder separator with serialized CTC to produce talker-ordered transcripts, and leverages an adapted LLM-based SOT objective as a \emph{multi-talker-aware} teacher signal to explicitly regularize mixed-speech representations.
To further support \emph{variable} numbers of talkers, we introduce a Talker-Count Head that predicts the talker count and dynamically selects the appropriate decoding branch. 
Experiments on LibriMix show that the proposed encoder-only model achieves comparable performance to LLM-based systems in the two-talker condition, while delivering significant improvements in the three-talker condition with significant small RTF. 
\end{abstract}

\section{Introduction}
\label{sec:intro}
Multi-talker automatic speech recognition (MT-ASR) \cite{shi2024advancing,shi_slt2024,meng2024large,8682822,8461893} aims to transcribe all talkers’ utterances from overlapping speech \cite{10542371,9689650,10038197,song22e_interspeech,zhao2024adapting}. 
End-to-end MT-ASR systems are often categorized by training strategy, where two representative lines are utterance-level permutation invariant training (uPIT) \cite{7979557,7952154} and serialized output training (SOT) \cite{kanda20b_interspeech}.
uPIT optimizes over all output permutations and thus scales poorly with the number of talkers, while typically assuming a fixed talker count.
SOT mitigates these issues by representing overlapped speech as a single serialized token sequence ordered by speaker onset time, and is commonly implemented with an attention-based encoder--decoder (AED) architecture \cite{6af3452a28a04980b2b8f5eb48730d36,chan2015listen}.

Recent progress in SOT-based MT-ASR has been largely decoder-centric, including improved labeling strategies \cite{shen2023speakermask}, enhanced self-attention \cite{10446059}, and adopting large language models (LLMs) as decoders \cite{shi2024advancing,meng2024large,shi2025serialized}.
However, relying on powerful decoders keeps the encoder largely talker-agnostic and forces the decoder to disentangle heavily overlapped representations over long sequences.
This is particularly suboptimal for LLM decoders: although they provide strong semantic modeling, they are slow and their gains do not reliably transfer to challenging three-talker mixtures, suggesting that the \emph{mixed-speech encoder representation} remains a key bottleneck. 
Motivated by this, recent encoder-side talker-related CTC approaches \cite{shi_slt2024,shi2025serialized,sakuma2025speakerdistinguishable} shift part of the disentanglement from the decoder to the encoder by introducing post-encoder modules such as a separator and talker-aware CTC heads, enabling faster encoder-only decoding.
Nevertheless, training serialized CTC under severe overlap can still be unstable without strong semantic regularization, and most CTC-based MT-ASR methods require assuming a fixed number of talkers \emph{a priori}.

In this paper, we shift the role of LLMs from an inference-time decoder to a \emph{train-time adaptable teacher} for encoder-only MT-ASR.
We first perform \emph{multi-talker information adaptation} on the LLM so that it better consumes and interprets talker-related cues under overlap, and we \emph{simultaneously distill} its semantic guidance into the encoder via an SOT-based objective to regularize mixed-speech representations.
Then, when training the post-encoder separator and serialized CTC heads, we \emph{continue} to use the adapted LLM as a teacher signal to preserve the encoder’s modeling capacity and stabilize optimization, while keeping inference fully encoder-only and CTC-style.
To support \emph{variable} talker numbers, we further introduce a Talker-Count Head (TCH) that predicts the number of talkers and dynamically routes inference to the corresponding two- or three-talker branch, mitigating the fixed-count limitation shared by prior talker-related CTC methods \cite{shi_slt2024,shi2025serialized,sakuma2025speakerdistinguishable}.


\section{Preliminaries}
\label{sec:prelim}

\subsection{Serialized Output Training (SOT) for MT-ASR}
\vspace{-5pt}
SOT generates a unified transcription by ordering the outputs of multiple talkers sequentially, based on the speaking start time of each talker.
To indicate a change of talker, a special token $\langle \mathrm{sc}\rangle$ is inserted between the transcriptions of different talkers.
For instance, with two talkers, the target sequence becomes
$\mathbf{T}_{\mathrm{sot}}=\{t_1^{1}, \dots, t_1^{N^{1}}, \langle \mathrm{sc}\rangle, t_2^{1}, \dots, t_2^{N^{2}}\}$,
where $t_1$ and $t_2$ correspond to the transcriptions of the first- and second-speaking talkers.
$N^{1}$ and $N^{2}$ correspond to the transcription lengths of the two talkers. 
Given this training target, the attention mechanism is able to attend to the relevant segments of overlapping speech encodings and decode the transcriptions $\mathbf{T}_e$ of multiple talkers in chronological order of their speaking onsets.
For SOT-based ASR, the training loss can be expressed as
\begin{equation}
\mathcal{L}_{\text{SOT}} = \mathcal{L}(\mathbf{T}_e, \mathbf{T}_{\mathrm{sot}}).
\label{eq:sot_ce}
\end{equation}
Since the encoder produces a unified embedding rather than explicitly separating overlapping speech, it is challenging to align the speech embeddings with the target label sequence through CTC \cite{shi_slt2024}.
Consequently, training of the SOT-based ASR system relies solely on cross-entropy (CE) loss.

\subsection{LLMs as Decoder for ASR}
\vspace{-5pt}
LLM-based ASR systems typically comprise three components: a speech encoder, a projector, and an LLM decoder.
Self-supervised learning (SSL) encoders \cite{9814838,10096630} are commonly adopted as the speech encoder.
Given an input waveform $\mathbf{y}$, the encoder produces an acoustic representation
\begin{equation}
\mathbf{H}_e = \operatorname{Encoder}(\mathbf{y}).
\end{equation}
where $\mathbf{H}_e$ denotes the speech encoding, which is usually much longer (in time) than the target text sequence; directly conditioning an LLM on it is computationally demanding.
A projector is therefore introduced to \emph{(i)} reduce the temporal resolution and \emph{(ii)} map the features into the text-embedding space:
\begin{equation}
\mathbf{H}_p = \operatorname{Projector}(\mathbf{H}_e).
\end{equation}
where $\mathbf{H}_p$ is the projected representation.
The projector consists of downsampling blocks followed by a dimensionality-mapping module.
Two common downsampling strategies are: (a) a stack of strided 2D convolutional layers, or (b) frame stacking that concatenates $n$ consecutive frames along the feature axis.
After downsampling, one or more linear layers project the features to the text-embedding space. 
The LLM decoder then generates the transcription conditioned on $\mathbf{H}_p$:
\begin{equation}
\mathbf{T}_e = \operatorname{LLM}(\mathbf{H}_p, \mathbf{E}_t).
\end{equation}
where $\mathbf{E}_t$ denotes the text embeddings and $\mathbf{T}_e$ is the decoded text.
How $\mathbf{H}_p$ is consumed depends on the LLM family: with BART-style decoders, $\mathbf{H}_p$ serves as encoder hidden states for cross-attention; with LLaMA-style decoders, $\mathbf{H}_p$ is typically concatenated (e.g., prepended or interleaved) with token embeddings at the input. 
During fine-tuning, the LLM is often kept fixed and only lightweight adapters (e.g., LoRA \cite{hu2021lora}) are updated.
Training employs the CE loss
\begin{equation}
\mathcal{L} = \operatorname{CE}(\mathbf{T}_l, \mathbf{T}_e).
\end{equation}
where $\mathbf{T}_l$ denotes the reference transcription.

\subsection{Serialized CTC}
\vspace{-5pt}
We introduce a \emph{post-encoder separator} that disentangles the mixed representation $\mathbf{H}_e$ into $S$ talker-specific streams ordered by their speaking onsets:
\begin{equation}
\mathbf{H}_{\text{sep}}^{1},\,\dots,\,\mathbf{H}_{\text{sep}}^{S}
= \operatorname{Separator}(\mathbf{H}_{e}).
\end{equation}
where $S$ is the number of talkers.
We instantiate the separator with an LSTM \cite{graves2012long} followed by layer normalization and $S$ parallel projection heads (one per talker):
\begin{equation}
\widetilde{\mathbf{H}} = \operatorname{LayerNorm}(\operatorname{LSTM}(\mathbf{H}_e)).
\end{equation}

Each talker-specific stream is then obtained via a linear head and nonlinearity:
\begin{equation}
\mathbf{H}_{\text{sep}}^{s}
= \operatorname{ReLU}(\operatorname{Linear}^{(s)}(\widetilde{\mathbf{H}})),
\quad s=1,\dots,S.
\end{equation}
The $s$-th head corresponds to the $s$-th talker in the SOT serialization order (i.e., by onset time).
Serialized CTC is applied independently to each stream:
\begin{equation}
\mathcal{L}_{\text{Serialized-CTC}}
= \sum_{s=1}^{S} \operatorname{Loss}_{\text{CTC}}\big(\mathbf{H}_{\text{sep}}^{s},\,\mathbf{T}^{s}\big).
\label{serilized_ctc_loss}
\end{equation}
where $\mathbf{T}^{s}$ is the transcription of the $s$-th talker in serialized order.

\begin{figure*}

    \centering
    \includegraphics[width=0.9\linewidth]{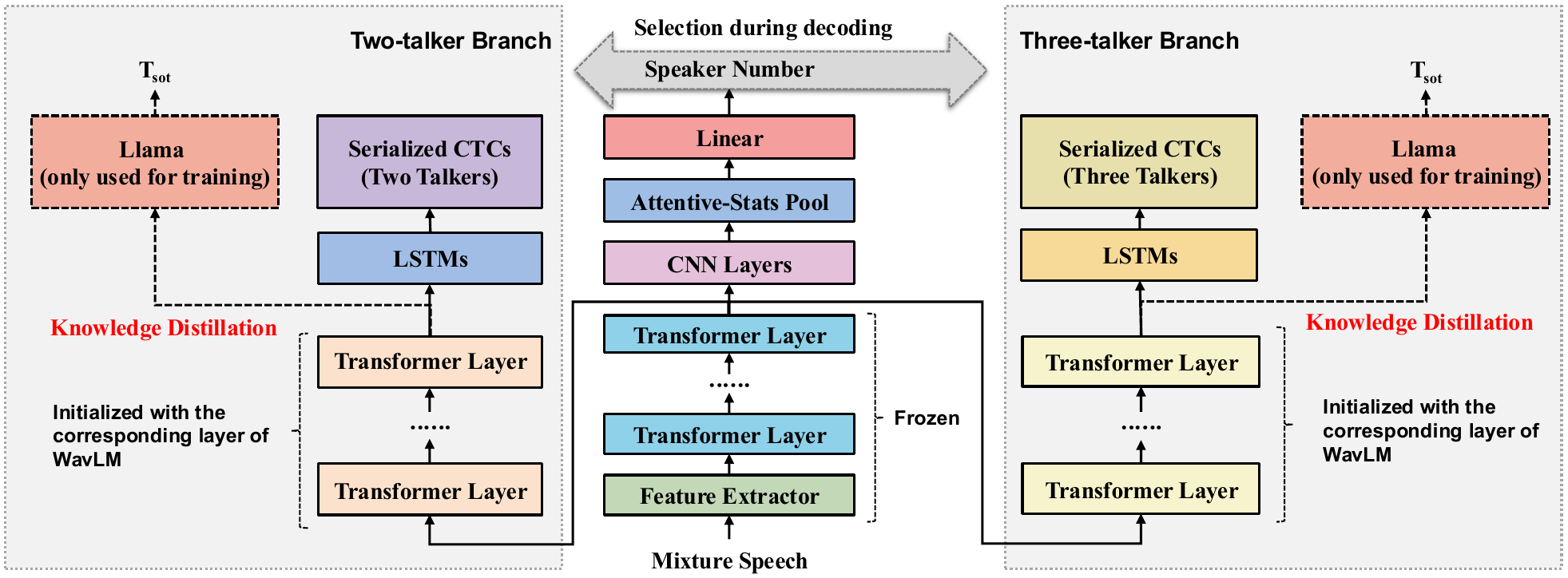}
    \vspace{-5pt}
    \caption{Flowchart of the proposed method. LLaMA is used only during training for distillation. 
    }
    \vspace{-15pt}
    \label{fig:architecture}
\end{figure*}

\section{Proposed Method}
\label{sec:proposed}
Our goal is to retain the efficiency of encoder-only MT-ASR while injecting semantic priors that are typically provided by autoregressive LLM decoders.
To this end, we propose an encoder-only model trained with a hybrid objective that distills LLM knowledge into the encoder via an SOT-based teacher signal, together with serialized CTC for fast inference.
In addition, since CTC-based MT-ASR commonly assumes a fixed number of talkers \emph{a priori}, we introduce a Talker-Count Head (TCH) that predicts the talker count and dynamically routes inference to the appropriate two- or three-talker branch.
The proposed model contains several shared encoding layers together with specialized two- and three-talker processing branches.

\subsection{Transcription Encoder with Serialized CTC}
\vspace{-5pt}
We adopt WavLM \cite{9814838} as the pretrained backbone of the transcription encoder.
Given an input waveform $\mathbf{y}$, a feature extractor $\mathrm{FE}$ produces frame-level features $\mathbf{F}=\mathrm{FE}(\mathbf{y})$.
A stack of shared Transformer layers $E_{\theta}$ further encodes them:
\begin{equation}
\mathbf{H}_s = E_{\theta}(\mathbf{F}).
\end{equation}
Two specialized Transformer branches, $B^{(2)}$ (two talkers) and $B^{(3)}$ (three talkers), refine $\mathbf{H}_s$:
\begin{equation}
\mathbf{Z}^{(b)} = B^{(b)}(\mathbf{H}_s)
= [\mathbf{z}^{(b)}_1,\ldots,\mathbf{z}^{(b)}_{T_b}],
\quad b\in\{2,3\}.
\end{equation}
Each branch is followed by a talker-wise separator that produces $b$ streams:
\begin{equation}
\mathbf{H}_{\mathrm{sep}}^{1},\,\ldots,\,\mathbf{H}_{\mathrm{sep}}^{b}
= \mathrm{Separator}^{(b)}(\mathbf{Z}^{(b)}),
\quad b\in\{2,3\}.
\end{equation}
The serialized CTC loss is then applied as in Eq.~(\ref{serilized_ctc_loss}).

\subsection{LLM Adaptation and Distillation}
\vspace{-5pt}
\label{sec:llm_distill}
Training serialized CTC under heavy overlap can be unstable without additional semantic regularization, and using an LLM as an inference-time decoder is costly.
We therefore use an LLM as a \emph{train-time adaptable teacher} and distill its semantic guidance into the encoder, while keeping inference fully encoder-only and CTC-style.

\textbf{Phase 1: Multi-talker information adaptation of the LLM with joint encoder distillation.}
We start from an SOT-based encoder--decoder model and use a pretrained LLaMA decoder.
To make the teacher better aligned with multi-talker conditioning, we adapt the LLM by updating only lightweight parameters (LoRA adapters \cite{hu2021lora} and token embeddings tied to the LM head to accommodate special tokens), while keeping the LLM backbone frozen.
This stage optimizes the SOT objective (Eq.~\ref{eq:sot_ce}), which serves two purposes:
(i) it adapts the LLM to better interpret talker-related cues under overlap, and
(ii) it \emph{simultaneously distills} the resulting semantic guidance into the encoder through backpropagation to the acoustic pathway.

\textbf{Phase 2: Serialized CTC training with continued distillation.}
We then attach the separator and serialized CTC heads after the corresponding two- or three-talker branch and optimize the encoder-only path.
During this phase, we keep the adapted LLaMA fixed and continue to compute the SOT loss as a teacher signal, so that the encoder maintains the semantic regularization learned in Phase 1 while being optimized for CTC decoding.
Concretely, we optimize a CTC--attention hybrid objective:
\begin{equation}
\mathcal{L}_{\text{EncSep}}
= \alpha\,\mathcal{L}_{\text{Serialized-CTC}} + (1-\alpha)\,\mathcal{L}_{\text{SOT}},
\label{eq:encsep}
\end{equation}
where $\alpha\in[0,1]$ balances the two terms.
Note that LLaMA is used only during training and introduces no additional computation at inference time.

\subsection{Talker-Count Head}
\vspace{-5pt}
The talker-count head maps the shared-encoder output $\mathbf{H}_s$ to binary logits $\mathbf{o}$ (two vs.\ three talkers).
We score each frame with an additive attention and apply an optional speech-activity mask $m_t\in\{0,1\}$.
Let $\mathbf{H}_s=[\mathbf{h}_1,\ldots,\mathbf{h}_T]$.
We compute attention weights as
\begin{equation}
\alpha_t
=
\frac{
\exp\bigl(\mathbf{v}^\top \tanh(\mathbf{W}\mathbf{h}_t+\mathbf{b}) + c\bigr)\,\mathbb{I}_{[m_t=1]}
}{
\sum_{t':\,m_{t'}=1}
\exp\bigl(\mathbf{v}^\top \tanh(\mathbf{W}\mathbf{h}_{t'}+\mathbf{b}) + c\bigr)
}.
\label{eq:count_attn}
\end{equation}
Using the attention weights $\alpha_t$, we compute mean and dispersion statistics:
\begin{equation}
\begin{aligned}
\boldsymbol{\mu} &= \sum_{t}\alpha_t\,\mathbf{h}_t, \\
\boldsymbol{\sigma} &=
\sqrt{\sum_{t}\alpha_t\left(\mathbf{h}_t-\boldsymbol{\mu}\right)^{\odot 2}+\varepsilon}, \\
\mathbf{z} &= [\boldsymbol{\mu};\boldsymbol{\sigma}] \in \mathbb{R}^{2D},
\end{aligned}
\label{eq:count_pool}
\end{equation}
where $D$ is the hidden dimensionality and $\odot$ denotes element-wise square.
The pooled vector is normalized and passed through a light MLP to obtain logits:
\begin{equation}
\mathbf{o}
=
\mathbf{W}_2\,\operatorname{Dropout}\bigl(
\operatorname{GELU}\bigl(\mathbf{W}_1\,\operatorname{LayerNorm}(\mathbf{z})+\mathbf{b}_1\bigr)
\bigr)
+\mathbf{b}_2.
\label{eq:count_logits}
\end{equation}
Class probabilities and the prediction are
\begin{equation}
\begin{aligned}
p_\theta(y{=}k \mid \mathbf{H}_s) &= \mathrm{softmax}(\mathbf{o})_k,\quad k\in\{1,2\},\\
\hat{y} &= \arg\max_k\, p_\theta(y{=}k \mid \mathbf{H}_s).
\end{aligned}
\label{eq:count_pred}
\end{equation}
We minimize CE loss over a minibatch of $N$ utterances:
\begin{equation}
\mathcal{L}_{\mathrm{count}}
=
-\frac{1}{N}\sum_{i=1}^{N}
\log
\frac{\exp\bigl(o^{(i)}_{y^{(i)}}\bigr)}
{\sum_{k=1}^{2}\exp\bigl(o^{(i)}_{k}\bigr)}.
\label{eq:count_loss}
\end{equation}
where $o^{(i)}_{y^{(i)}}$ denotes the logit assigned by the model to the ground-truth class $y^{(i)}$ for sample $i$ (i.e., the pre-softmax score).

\section{Experiments}
\label{sec:exp}
\subsection{Datasets}
\vspace{-5pt}
We evaluate our models on LibriMix \cite{cosentino2020librimix}. 
Clean speech is from the LibriSpeech \texttt{train-clean-100}, \texttt{train-clean-360}, \texttt{dev-clean}, and \texttt{test-clean} subsets \cite{7178964}, and noise samples are taken from WHAM! \cite{Wichern2019WHAM}. 
Two- and three-talker mixtures are synthesized using the official scripts. 
To control overlap patterns, we use start-time \emph{offset files}: for two-talker mixtures we follow the official ESPnet configuration, whereas for three-talker mixtures we create our own offsets.

\subsection{Model Configurations}
\vspace{-5pt}
We initialize the encoder stacks from the pretrained WavLM-Large (24 Transformer blocks)\footnote{\scriptsize\url{https://huggingface.co/microsoft/wavlm-large}}, owing to its robust pretraining, including mixture/overlap simulation well suited to multi-talker conditions. 
For example, the shared encoder (12 layers) is seeded with WavLM layers 1–12, while the two- and three-talker branches (12 layers each) are initialized from layers 13–24 (independent copies, no weight tying). 
During training, the shared layers are frozen. 
For the LLM decoder, we use LLaMA-3.2-1B\footnote{\scriptsize\url{https://huggingface.co/meta-llama/Llama-3.2-1B}}. 
For Serialized CTC, the Separator contains the two-layer LSTM with 896 hidden units.

\begin{table*}[t]
\caption{Results of the proposed method. 
``S/V'' denotes the talker-count condition used in training:
S = single-count (only one talker number), V = varied-count (mixed two-/three-talker). 
``FE'' is the feature extraction module. 
``TCH'' represents the talker-count head. 
``TCH.L.'' represents the number of TCH layers. 
``Enc.L.'' represents the number of each encoder layers for transcription. 
``A.TC'' is the talker-count accuracy (\%). 
Word Error Rate (WER) is used for 2mix and 3mix evaluation.}
\footnotesize
\vspace{-10pt}
\label{table:results}
\renewcommand{\arraystretch}{1.}
\centering
{
\setlength{\tabcolsep}{3pt}
\renewcommand{\arraystretch}{0.95}

\begin{tabular}{@{}
  c|c|c|c|c|c|
  C{7mm}C{7mm}C{7mm}|
  C{7mm}C{7mm}C{7mm}|
  C{7mm}C{7mm}C{7mm}|
  C{7mm}C{7mm}C{7mm}
@{}}
\toprule[1.5pt]
\multirow{3}{*}{\textbf{ID}}
& \multirow{3}{*}{\textbf{S/V}} & \multirow{3}{*}{\textbf{Model}}
& \multicolumn{3}{c|}{\textbf{Encoder}}
& \multicolumn{6}{c|}{\textbf{Noisy}}
& \multicolumn{6}{c}{\textbf{Clean}} \\
\cline{4-18}
& & & \multirow{2}{*}{\textbf{FE}} & \multirow{2}{*}{\textbf{TCH.L.}} & \multirow{2}{*}{\textbf{Enc.L.}}
& \multicolumn{3}{c|}{\textbf{Development Set}}
& \multicolumn{3}{c|}{\textbf{Evaluation Set}}
& \multicolumn{3}{c|}{\textbf{Development Set}}
& \multicolumn{3}{c}{\textbf{Evaluation Set}} \\
\cline{7-18}
& & & & & &
\textbf{A.TC} & \textbf{2mix} & \textbf{3mix}
& \textbf{A.TC} & \textbf{2mix} & \textbf{3mix}
& \textbf{A.TC} & \textbf{2mix} & \textbf{3mix}
& \textbf{A.TC} & \textbf{2mix} & \textbf{3mix} \\
\midrule

1 & \multirow{4}{*}{S} & \multirow{4}{*}{\makecell[c]{(Baseline)\\SOT\\w/ Llama-1B}}
& \ding{51} & - & 24
& - & 12.4 & 39.8
& - & 11.3 & 39.1
& - & 4.6 & 21.5
& - & 4.6 & 21.6 \\

2 & & 
& \ding{51} & - & 18
& - & 17.7 & 49.2
& - & 15.5 & 48.4
& - & 5.9 & 30.4
& - & 6.1 & 31.7 \\

3 & & & \ding{51} &
- & 12
& - & 35.8 & 62.0
& - & 34.2 & 62.2
& - & 25.3 & 47.7
& - & 25.1 & 49.7 \\

4 & & & \ding{51} &
- & 6
& - & 43.1 & 65.5
& - & 41.2 & 65.4
& - & 29.6 & 52.5
& - & 30.5 & 54.7 \\

\midrule


\textcolor{gray!75}{5} & \multirow{6}{*}{V} & \multicolumn{3}{c|}{\textcolor{gray!75}{Training without KD from LLM}}
& \textcolor{gray!75}{24}
& \textcolor{gray!75}{-} & \textcolor{gray!75}{68.0} & \textcolor{gray!75}{81.6}
& \textcolor{gray!75}{-} & \textcolor{gray!75}{66.3} & \textcolor{gray!75}{80.4}
& \textcolor{gray!75}{-} & \textcolor{gray!75}{51.8} & \textcolor{gray!75}{69.9}
& \textcolor{gray!75}{-} & \textcolor{gray!75}{52.2} & \textcolor{gray!75}{70.1} \\

\cline{3-18}

6 & & \multirow{5}{*}{\makecell[c]{(Proposed)\\ Encoder-only \\ Model}}
& - & - & 24
& \hly{100.} & 10.5 & 23.3
& \hly{100.} & 9.6 & 22.2
& \hly{100.} & 4.0 & 12.8
& \hly{100.} & 4.1 & 12.9 \\

7 & &
& \ding{51} & - & 24
& 80.2 & 12.0 & 28.0
& 78.1 & 11.5 & 26.9
& 89.9 & 4.5 & 15.8
& 88.0 & 4.6 & 16.2 \\

8 & & & \ding{51} &
6 & 24
& 88.5 & 10.6 & 28.0
& 86.8 & 9.6 & 28.3
& 97.8 & 4.0 & 13.9
& 96.3 & 4.1 & 14.6 \\

9 & & & \ding{51} &
12 & 24
& 94.6 & 10.7 & 25.1
& 93.7 & 9.7 & 24.5
& 98.0 & 4.0 & 13.7
& 97.0 & 4.1 & 14.3 \\

10 & & & \ding{51} &
18 & 24
& 89.7 & 10.6 & 27.2
& 88.0 & 9.6 & 27.2
& 97.0 & 4.0 & 14.2
& 95.8 & 4.1 & 14.7 \\

\bottomrule[1.5pt]
\end{tabular}
}
\vspace{-15pt}
\end{table*}

\begin{table}[t]
\caption{Accuracy of TCH for 2mix and 3mix condition}
\footnotesize
\vspace{-10pt}
\label{table:tch}
\renewcommand{\arraystretch}{1.1}
\centering
{
\setlength{\tabcolsep}{3pt}
\renewcommand{\arraystretch}{0.95}

\begin{tabular}{@{}
  c|c|
  C{6mm}C{6mm}|
  C{6mm}C{6mm}|
  C{6mm}C{6mm}|
  C{6mm}C{6mm}
@{}}
\toprule[1.5pt]
\multirow{3}{*}{\textbf{ID}}
& \multirow{3}{*}{\textbf{TCH.L.}}
& \multicolumn{4}{c|}{\textbf{Noisy}}
& \multicolumn{4}{c}{\textbf{Clean}} \\
\cline{3-10}
& 
& \multicolumn{2}{c|}{\textbf{Dev.}}
& \multicolumn{2}{c|}{\textbf{Eval.}}
& \multicolumn{2}{c|}{\textbf{Dev.}}
& \multicolumn{2}{c}{\textbf{Eval.}} \\
\cline{3-10}
& &
 \textbf{2mix} & \textbf{3mix}
& \textbf{2mix} & \textbf{3mix}
 & \textbf{2mix} & \textbf{3mix}
 & \textbf{2mix} & \textbf{3mix} \\
\midrule

7 
& - 
& 78.8 & 81.7
& 73.4 & 82.8
& 90.9 & 89.0
& 88.8 & 87.3 \\

8 
& 6 
& 99.7 & 77.3
& 99.9 & 73.7
& 100. & 95.5
& 99.9 & 92.6 \\

9 & 12 
& 99.0 & 90.2
& 98.7 & 88.7
& 100. & 96.0
& 99.8 & 94.1 \\

10 & 18
& 99.7 & 79.7
& 99.3 & 76.7
& 99.9 & 94.0
& 99.9 & 91.7 \\

\bottomrule[1.5pt]
\end{tabular}
}
\vspace{-10pt}
\end{table}

\begin{table}[]
\centering
\caption{Real-time factor (RTF) between the proposed methods and SOT-based LLM methods.}
\vspace{-10pt}
\begin{tabular}{l|cc}
\toprule[1.5pt]
Model    & Libri2Mix & Libri3Mix \\
\midrule
CTC      & 0.0043    & 0.0106    \\
Llama-1B & 0.1150    & 0.0981   \\ 
\bottomrule[1.5pt]
\end{tabular}
\vspace{-15pt}
\label{table:rtf}
\end{table}

\subsection{Experimental Results and Analysis}
\vspace{-5pt}
\subsubsection{Impact of Encoder Depth}
\vspace{-5pt}
We first evaluate baseline models to examine how the number of encoder layers influences ASR performance, as shown in Table~\ref{table:results}. 
From ID--1 to ID--4, the results indicate that recognition performance drops substantially as the number of encoder layers decreases. 
Thus, we use all 24 Transformer encoder layers of WavLM-Large for the encoder. The two- and three-talker encoders share only the feature extraction module of WavLM.

\subsubsection{Performance of the Encoder-Only Model}
\vspace{-5pt}
We first examined training the model with only the serialized CTC loss (Eq.~\ref{serilized_ctc_loss}) in ID--5. 
The results show that the model fails to train effectively under this setting. 
We evaluated the encoder-only model under the assumption of a perfectly accurate talker number classifier. 
As shown in ID--6 in Table~\ref{table:results}, the proposed encoder-only model achieved strong performance, with significant improvements over ID--1 across all conditions, especially in the 3-mix scenario. 
We then evaluated the encoder-only model under different configurations of the Talker-Count Head (TCH). 
In ID--7, only the feature extraction module is used prior to the TCH. 
In ID--8 to ID--10, the feature extractor is followed by varying numbers of encoder layers (initialized from the corresponding layers of WavLM) before the TCH. 
During training, both the feature extractor and all Transformer layers are kept frozen. 
The results suggest that adding more Transformer layers does not always lead to better performance. 
Considering all conditions, ID--9 (with 12 Transformer layers before the TCH) yields the most effective improvement among the tested settings (0/6/12/18 layers). 
For the 2-talker condition, all TCH configurations with Transformer layers yielded similar performance. 
For the 3-talker condition, although adding 6 or 18 Transformer layers improved talker-count accuracy compared with using only the feature extraction module, the ASR performance did not improve under noisy condition. 
Thus, we further examined the detailed accuracy of each TCH setting to identify the underlying reasons, which is shown in Table~\ref{table:tch}. 
The results show that with Transformer layers, the 2-talker condition can be recognized with very high accuracy, whereas the 3-talker condition remains much more difficult to achieve high accuracy. 
While ID--9 achieves substantial gains compared with the baseline model ID--1 under the 3-talker condition, it still performs worse than ID--6. 
The claim that preconditioning with accurate talker counting is an essential module for MT-ASR \cite{cornell24_chime, Cornell2025RecentTrends} is also supported by the conclusions of this paper, even without additional experiments. 

\subsubsection{Inference Efficiency of the Encoder-Only Model}
\vspace{-5pt}
We reported the RTF for the proposed CTC-based model and the SOT-based LLM baseline on Libri2Mix and Libri3Mix. 
RTF was computed as the ratio of total inference time to total input audio duration, measured on a single NVIDIA H100 80GB GPU with batch size 1. 
As shown in Table \ref{table:rtf}, the CTC model ran substantially faster than Llama-1B.

\subsubsection{Comparison between existing approaches}
\vspace{-5pt}
Table~\ref{table:method_comparision} compares the proposed method with existing approaches. 
The results show that larger LLMs can improve performance in the 2-talker condition, but our proposed encoder-only method achieves comparable results. 
While LLM-based methods struggle with the 3-talker condition as analyzed in our previous work \cite{shi2025serialized}, the encoder-only method performs well and even outperforms the LLM-based approach. 
We also compare a variant without an LLM, using a standard Transformer decoder (\textit{w/o LLM}), as well as encoder-only CTC decoding (\textit{w/o decoder: CTC}). 
Results show that stronger semantic guidance from the LLM consistently improves performance, and remains beneficial even in the three-talker condition, indicating that semantic information significantly strengthens the encoder under severe overlap.

\begin{table}[!h]
\renewcommand{\arraystretch}{1.}
\caption{Comparison between the proposed method and existing methods on the LibriMix datasets (270 hours for Libri2Mix and 186 hours for Libri3Mix, without any additional data augmentation). Word Error Rate (WER) is used for evaluation. ``N/C'' represents the noisy or clean condition.}
\footnotesize
\vspace{-10pt}
\centering
\begin{tabular}{ll|cc|cc}
\toprule[1.5pt]
\multicolumn{2}{c|}{\multirow{2}{*}{\textbf{REF}}}
& \multicolumn{2}{c|}{\textbf{Libri2Mix}} & \multicolumn{2}{c}{\textbf{Libri3Mix}} \\
\cline{3-6}
& & \textbf{Dev} & \textbf{Eval} & \textbf{Dev} & \textbf{Eval} \\

\midrule

\multicolumn{1}{c|}{\multirow{8}{*}{\textbf{N}}} & \multicolumn{5}{c}{\textbf{Without LLMs; with SSL for the speech encoder}} \\
\cline{2-6}

\multicolumn{1}{c|}{} & Training from Scratch \tablefootnote{\scriptsize\url{https://github.com/espnet/espnet/tree/master/egs2/librimix/sot_asr1} \label{myfoot}} & 19.4 & 17.1 & 30.5 & 28.2 \\




\multicolumn{1}{c|}{} & TSE-V-Whisper \cite{10389752} & - & 12.0 & - & - \\

\multicolumn{1}{c|}{} & GEncSep \cite{shi_slt2024} & 17.2 & 15.0 & 28.0 & 25.9 \\

\multicolumn{1}{c|}{} & \quad w/o decoder: CTC & 16.8 & 14.6 & 25.7 & 23.6 \\

\cline{2-6}

\multicolumn{1}{c|}{} & \multicolumn{5}{c}{\textbf{With LLMs}} \\
\cline{2-6}

\multicolumn{1}{c|}{} & SOT-Llama-1B \cite{shi2025serialized} & 12.4 & 11.3 & 39.8 & 39.1 \\

\multicolumn{1}{c|}{} & SOP-Llama-1B \cite{shi2025serialized} & 11.8 & 10.5 & 29.6 & 28.5 \\

\multicolumn{1}{c|}{} & SOT-Llama-3B \cite{shi2025serialized} & 11.2 & 9.8 & 34.2 & 31.7 \\

\multicolumn{1}{c|}{} & SOP-Llama-3B \cite{shi2025serialized} & {10.5} & {9.2} & {29.3} & {28.1} \\
\cline{2-6}

\multicolumn{1}{c|}{} & ID-8 & 10.7 & 9.7 & 25.1 & 24.5 \\

\midrule

\multicolumn{1}{c|}{\multirow{9}{*}{\textbf{C}}} & \multicolumn{5}{c}{\textbf{Without LLMs; with SSL for the speech encoder}} \\
\cline{2-6}

\multicolumn{1}{c|}{} & Training from Scratch \footnotemark[\value{footnote}] & 6.8 & 7.0 & 15.0 & 14.7 \\


\multicolumn{1}{c|}{} & W2V-Sidecar-ft. \cite{10095295} & 7.7 & 8.1 & - & - \\

\multicolumn{1}{c|}{} & WavLM-CLN \cite{10097139} & 7.1 & 7.6 & - & - \\

\multicolumn{1}{c|}{} & C-HuBERT LARGE \cite{10096630} & 6.6 & 7.8 & - & - \\


\multicolumn{1}{c|}{} & GEncSep \cite{shi_slt2024} & 6.4 & 6.6 & 13.3 & 13.1 \\

\multicolumn{1}{c|}{} & \quad w/o decoder: CTC & 6.0 & 6.1 & 12.7 & 12.5 \\

\cline{2-6}

\multicolumn{1}{c|}{} & \multicolumn{5}{c}{\textbf{With LLMs}} \\
\cline{2-6}

\multicolumn{1}{c|}{} & SOT-Llama-1B \cite{shi2025serialized} & 4.6 & 4.6 & 21.5 & 21.6 \\

\multicolumn{1}{c|}{} & SOP-Llama-1B \cite{shi2025serialized} & 3.9 & 4.0 & 20.8 & 22.0 \\

\multicolumn{1}{c|}{} & SOT-Llama-3B \cite{shi2025serialized} & 4.0 & 4.1 & 22.3 & 22.0 \\

\multicolumn{1}{c|}{} & SOP-Llama-3B \cite{shi2025serialized} & {3.5}	& {3.6} & {17.0} & {16.5} \\
\cline{2-6}
\multicolumn{1}{c|}{} & ID-8 & 4.0 & 4.1 & 13.7 & 14.3 \\

\bottomrule[1.5pt]

\end{tabular}
\vspace{-15pt}
\label{table:method_comparision}
\end{table}

\section{Conclusions}
\label{sec:conclu}
In this paper, we proposed an encoder-only framework for multi-talker ASR that preserves the decoding efficiency of serialized CTC while injecting semantic priors from large language models (LLMs) during training.
Specifically, we adapt an LLM to multi-talker conditioning and distill its guidance into the encoder, and we further continue the distillation signal when optimizing the post-encoder separator and serialized CTC heads, yielding a fast CTC-style inference pipeline without relying on an LLM decoder at test time.
To support variable numbers of talkers, we introduced a Talker-Count Head (TCH) that dynamically selects the appropriate two- or three-talker branch, removing the need to pre-specify the talker count. 
Experiments on Libri2Mix and Libri3Mix showed that our encoder-only model achieved performance comparable to LLM-based systems in the two-talker setting, while providing substantial gains in the more challenging three-talker condition.
Although TCH estimation remained highly accurate for two-talker mixtures and was less reliable for three-talker mixtures, TCH-guided routing still led to consistent improvements and enabled the overall system to outperform LLM-based approaches in Libri3Mix.
Future work will focus on improving talker-count robustness under severe overlap and noise, and extending the framework to broader variable-talker conditions.

\section{Generative AI Use Disclosure}
Generative AI tools (Gemini and ChatGPT) were used for language editing and improving the phrasing of this manuscript.

\bibliographystyle{IEEEtran}
\bibliography{mybib}

@ARTICLE{9814838,
  author={Chen, Sanyuan and Wang, Chengyi and Chen, Zhengyang and Wu, Yu and Liu, Shujie and Chen, Zhuo and Li, Jinyu and Kanda, Naoyuki and Yoshioka, Takuya and Xiao, Xiong and Wu, Jian and Zhou, Long and Ren, Shuo and Qian, Yanmin and Qian, Yao and Wu, Jian and Zeng, Michael and Yu, Xiangzhan and Wei, Furu},
  journal={IEEE JSTSP}, 
  title={WavLM: Large-Scale Self-Supervised Pre-Training for Full Stack Speech Processing}, 
  year={2022},
  volume={16},
  number={6},
  pages={1505-1518},
  doi={10.1109/JSTSP.2022.3188113}}

@INPROCEEDINGS{10096630,
  author={Fazel-Zarandi, Maryam and Hsu, Wei-Ning},
  booktitle={Proc. ICASSP}, 
  title={Cocktail Hubert: Generalized Self-Supervised Pre-Training for Mixture and Single-Source Speech}, 
  year={2023},
  volume={},
  number={},
  pages={1-5},
  doi={10.1109/ICASSP49357.2023.10096630}}

@ARTICLE{shi2025serialized,
  author    = {Hao Shi and Yusuke Fujita and Tomoya Mizumoto and Lianbo Liu and Atsushi Kojima and Yui Sudo},
  title     = {Serialized Output Prompting for Large Language Model-based Multi-Talker Speech Recognition},
  journal   = {arXiv preprint arXiv:2509.04488},
  year      = {2025},
}

@ARTICLE{10542371,
  author={Shi, Hao and Mimura, Masato and Kawahara, Tatsuya},
  journal={IEEE/ACM TASLP}, 
  title={Waveform-Domain Speech Enhancement Using Spectrogram Encoding for Robust Speech Recognition}, 
  year={2024},
  volume={32},
  number={},
  pages={3049-3060},
  doi={10.1109/TASLP.2024.3407511}}

@inproceedings{song22e_interspeech,
  author={Tongtong Song and Qiang Xu and Meng Ge and Longbiao Wang and Hao Shi and Yongjie Lv and Yuqin Lin and Jianwu Dang},
  title={{Language-specific Characteristic Assistance for Code-switching Speech Recognition}},
  year=2022,
  booktitle={Proc. Interspeech},
  pages={3924--3928},
  doi={10.21437/Interspeech.2022-11426}
}

@INPROCEEDINGS{shi2024advancing,
  author={Shi, Mohan and Jin, Zengrui and Xu, Yaoxun and Xu, Yong and Zhang, Shi-Xiong and Wei, Kun and Shao, Yiwen and Zhang, Chunlei and Yu, Dong},
  booktitle={Proc. SLT}, 
  title={Advancing Multi-Talker {ASR} Performance With Large Language Models}, 
  year={2024},
  volume={},
  number={},
  pages={14-21},
}

@INPROCEEDINGS{shi_slt2024,
  author={Shi, Hao and Gao, Yuan and Ni, Zhaoheng and Kawahara, Tatsuya},
  booktitle={Proc. SLT}, 
  title={Serialized Speech Information Guidance with Overlapped Encoding Separation for Multi-Speaker Automatic Speech Recognition}, 
  year={2024},
  volume={},
  number={},
  pages={198--204},
}

@INPROCEEDINGS{meng2024large,
  author={Meng, Lingwei and Hu, Shujie and Kang, Jiawen and Li, Zhaoqing and Wang, Yuejiao and Wu, Wenxuan and Wu, Xixin and Liu, Xunying and Meng, Helen},
  booktitle={Proc. ICASSP}, 
  title={Large Language Model Can Transcribe Speech in Multi-Talker Scenarios with Versatile Instructions}, 
  year={2025},
  volume={},
  number={},
  pages={1-5},
}

@INPROCEEDINGS{10038197,
  author={Yang, Yanbing and Shi, Hao and Lin, Yuqin and Ge, Meng and Wang, Longbiao and Hou, Qingzhi and Dang, Jianwu},
  booktitle={Proc. ISCSLP}, 
  title={Adaptive Attention Network with Domain Adversarial Training for Multi-Accent Speech Recognition}, 
  year={2022},
  volume={},
  number={},
  pages={6-10},
  doi={10.1109/ISCSLP57327.2022.10038197}}

@INPROCEEDINGS{zhao2024adapting,
  author={Zhao, Jiahui and Shi, Hao and Cui, Chenrui and Wang, Tianrui and Liu, Hexin and Ni, Zhaoheng and Ye, Lingxuan and Wang, Longbiao},
  booktitle={Proc. ICASSP}, 
  title={Adapting Whisper for Code-Switching through Encoding Refining and Language-Aware Decoding}, 
  year={2025},
  volume={},
  number={},
  pages={1-5},
}

@INPROCEEDINGS{9689650,
  author={Shi, Hao and Wang, Longbiao and Li, Sheng and Fan, Cunhang and Dang, Jianwu and Kawahara, Tatsuya},
  booktitle={Proc. APSIPA ASC}, 
  title={Spectrograms Fusion-based End-to-end Robust Automatic Speech Recognition}, 
  year={2021},
  volume={},
  number={},
  pages={438-442},
  doi={}}

@INPROCEEDINGS{8682822,
  author={Chang, Xuankai and Qian, Yanmin and Yu, Kai and Watanabe, Shinji},
  booktitle={Proc. ICASSP}, 
  title={End-to-end Monaural Multi-speaker {ASR} System without Pretraining}, 
  year={2019},
  volume={},
  number={},
  pages={6256-6260},
  doi={10.1109/ICASSP.2019.8682822}}

@INPROCEEDINGS{8461893,
  author={Settle, Shane and Roux, Jonathan Le and Hori, Takaaki and Watanabe, Shinji and Hershey, John R.},
  booktitle={Proc. ICASSP}, 
  title={End-to-End Multi-Speaker Speech Recognition}, 
  year={2018},
  volume={},
  number={},
  pages={4819-4823},
  doi={10.1109/ICASSP.2018.8461893}}

@ARTICLE{7979557,
  author={Kolbæk, Morten and Yu, Dong and Tan, Zheng-Hua and Jensen, Jesper},
  journal={IEEE/ACM TASLP}, 
  title={Multitalker Speech Separation With Utterance-Level Permutation Invariant Training of Deep Recurrent Neural Networks}, 
  year={2017},
  volume={25},
  number={10},
  pages={1901-1913},
  doi={10.1109/TASLP.2017.2726762}}

@INPROCEEDINGS{7952154,
  author={Yu, Dong and Kolbæk, Morten and Tan, Zheng-Hua and Jensen, Jesper},
  booktitle={Proc. ICASSP}, 
  title={Permutation invariant training of deep models for speaker-independent multi-talker speech separation}, 
  year={2017},
  volume={},
  number={},
  pages={241-245},
  doi={10.1109/ICASSP.2017.7952154}}

@inproceedings{kanda20b_interspeech,
  author={Kanda, Naoyuki and Gaur, Yashesh and Wang, Xiaofei and Meng, Zhong and Yoshioka, Takuya},
  title={{Serialized Output Training for End-to-End Overlapped Speech Recognition}},
  year=2020,
  booktitle={Proc. Interspeech},
  pages={2797--2801},
}

@inproceedings{6af3452a28a04980b2b8f5eb48730d36,
title = "{End-to-end continuous speech recognition using attention-based recurrent NN: First results}",
author = "Chorowski, Jan and Bahdanau, Dzmitry and Cho, Kyunghyun and Bengio, Yoshua",
year = "2014",
booktitle = "NIPS 2014 Workshop on Deep Learning",
}

@INPROCEEDINGS{chan2015listen,
  author={Chan, William and Jaitly, Navdeep and Le, Quoc and Vinyals, Oriol},
  booktitle={Proc. ICASSP}, 
  title={Listen, attend and spell: A neural network for large vocabulary conversational speech recognition}, 
  year={2016},
  volume={},
  number={},
  pages={4960-4964},
}

@article{hu2021lora,
  title={Lora: Low-rank adaptation of large language models},
  author={Hu, Edward J and Shen, Yelong and Wallis, Phillip and Allen-Zhu, Zeyuan and Li, Yuanzhi and Wang, Shean and Wang, Lu and Chen, Weizhu},
  journal={arXiv preprint arXiv:2106.09685},
  year={2021}
}

@article{cosentino2020librimix,
  title={Librimix: An open-source dataset for generalizable speech separation},
  author={Cosentino, Joris and Pariente, Manuel and Cornell, Samuele and Deleforge, Antoine and Vincent, Emmanuel},
  journal={arXiv preprint arXiv:2005.11262},
  year={2020}
}

@INPROCEEDINGS{7178964,
  author={Panayotov, Vassil and Chen, Guoguo and Povey, Daniel and Khudanpur, Sanjeev},
  booktitle={Proc. ICASSP}, 
  title={Librispeech: An {ASR} corpus based on public domain audio books}, 
  year={2015},
  volume={},
  number={},
  pages={5206-5210},
  doi={10.1109/ICASSP.2015.7178964}}

@inproceedings{Wichern2019WHAM,
    title     = {WHAM!: Extending Speech Separation to Noisy Environments},
    author    = {Wichern, Gordon and Antognini, Joe and Flynn, Michael and Zhu,
                 Licheng Richard and McQuinn, Emmett and Crow,
                 Dwight and Manilow, Ethan and Le Roux, Jonathan},
    booktitle = {Proc. Interspeech},
    year      = {2019},
    month     = sep
}

@article{graves2012long,
  title={Long short-term memory},
  author={Graves, Alex and Graves, Alex},
  journal={Supervised sequence labelling with recurrent neural networks},
  pages={37--45},
  year={2012},
  publisher={Springer}
}

@INPROCEEDINGS{10095295,
  author={Meng, L. and Kang, J. and Cui, M. and Wang, Y. and Wu, X. and Meng, H.},
  booktitle={Proc. ICASSP}, 
  title={A Sidecar Separator Can Convert A Single-Talker Speech Recognition System to A Multi-Talker One}, 
  year={2023},
  volume={},
  number={},
  pages={1-5},
}

@INPROCEEDINGS{10097139,
  author={Huang, Z. and Raj, D. and García, P. and Khudanpur, S.},
  booktitle={Proc. ICASSP}, 
  title={Adapting Self-Supervised Models to Multi-Talker Speech Recognition Using Speaker Embeddings}, 
  year={2023},
  volume={},
  number={},
  pages={1-5},
}

@INPROCEEDINGS{10389752,
  author={Zhang, W. and Yang, L. and Qian, Y.},
  booktitle={Proc. ASRU}, 
  title={Exploring Time-Frequency Domain Target Speaker Extraction For Causal and Non-Causal Processing}, 
  year={2023},
  volume={},
  number={},
  pages={1-6},
}

@ARTICLE{shen2023speakermask,
  author    = {P. Shen and X. Lu and H. Kawai},
  title     = {Speaker Mask Transformer for Multi-talker Overlapped Speech Recognition},
  journal   = {arXiv preprint arXiv:2312.10959},
  year      = {2023},
  url       = {https://arxiv.org/abs/2312.10959}
}

@INPROCEEDINGS{10446059,
  author={Fan, Zhiyun and Dong, Linhao and Zhang, Jun and Lu, Lu and Ma, Zejun},
  booktitle={Proc. ICASSP}, 
  title={SA-SOT: Speaker-Aware Serialized Output Training for Multi-Talker ASR}, 
  year={2024},
  volume={},
  number={},
  pages={9986-9990},
  doi={10.1109/ICASSP48485.2024.10446059}}

@ARTICLE{sakuma2025speakerdistinguishable,
  author    = {A. Sakuma and H. Sato and R. Sugano and T. Kumano and Y. Kawai and T. Ogawa},
  title     = {Speaker-Distinguishable CTC: Learning Speaker Distinction Using CTC for Multi-Talker Speech Recognition},
  journal   = {arXiv preprint arXiv:2506.07515},
  year      = {2025},
}

@inproceedings{cornell24_chime,
  title     = {The CHiME-8 DASR Challenge for Generalizable and Array Agnostic Distant Automatic Speech Recognition and Diarization},
  author    = {Samuele Cornell and Tae Jin Park and He Huang and Christoph Boeddeker and Xuankai Chang and Matthew Maciejewski and Matthew S Wiesner and Paola Garcia and Shinji Watanabe},
  year      = {2024},
  booktitle = {8th International Workshop on Speech Processing in Everyday Environments (CHiME 2024)},
  pages     = {1--6},
  doi       = {10.21437/CHiME.2024-1},
}

@ARTICLE{Cornell2025RecentTrends,
  author    = {S. Cornell and C. Boeddeker and T. Park and H. Huang and D. Raj and M. Wiesner and Y. Masuyama and X. Chang and Z.-Q. Wang and S. Squartini and P. Garcia and S. Watanabe},
  title     = {Recent Trends in Distant Conversational Speech Recognition: A Review of {CHiME-7} and 8 {DASR} Challenges},
  journal   = {arXiv preprint arXiv:2507.18161},
  year      = {2025},
  url       = {https://arxiv.org/abs/2507.18161}
}

\end{document}